\documentclass[lettersize,journal]{IEEEtran}
\usepackage{amsmath,amsfonts}
\usepackage{algorithmic}
\usepackage{algorithm}
\usepackage{array}
\usepackage[caption=false,font=normalsize,labelfont=sf,textfont=sf]{subfig}
\usepackage{textcomp}
\usepackage{stfloats}
\usepackage{url}
\usepackage{verbatim}
\usepackage{graphicx}
\usepackage{cite}
\usepackage{braket}
\usepackage{bm}
\usepackage{multirow}
\begin{document}

\title{Bayesian Quantum Neural Network for Renewable- Rich Power Flow with Training Efficiency and Generalization Capability Improvements}

\author{Ziqing Zhu,~\IEEEmembership{Member,~IEEE,} Shuyang Zhu,~\IEEEmembership{Student Member,~IEEE,} Siqi Bu,~\IEEEmembership{Senior Member,~IEEE} 
}
\maketitle

\begin{abstract}
This paper addresses the challenges of power flow calculation in large-scale power systems with high renewable penetration, focusing on computational efficiency and generalization. Traditional methods, while accurate, struggle with scalability for large power systems. Existing data-driven deep learning approaches, despite their speed, require extensive training data and lacks generalization capability in face of unseen scenarios, such as uncertainties of power flow caused by renewables. To overcome these limitations, we propose a novel power flow calculation model based on Bayesian Quantum Neural Networks (BQNNs). This model leverages quantum computing's ability to improve the training efficiency. The BQNN is trained using Bayesian methods, enabling it to update its understanding of renewable energy uncertainties dynamically, improving generalization to unseen data. Additionally, we introduce two evaluation metrics: effective dimension for model complexity and generalization error bound to assess the model’s performance in unseen scenarios. Our approach demonstrates improved training efficiency and better generalization capability, making it as an effective tool for future steady-state power system analysis.

\end{abstract}


\section{Introduction}
\IEEEPARstart{P}{ower} flow calculation, also known as load flow analysis, is a fundamental tool in the operation and planning of electrical power systems \cite{liu2010improved}. It involves determining the voltage, current, and power flows across all components of a grid under steady-state conditions \cite{tang2019robust}. This analysis helps ensure that the system operates within safe limits and identifies potential issues like voltage instability or line overloads, which could lead to blackouts \cite{wang2017three}. 

Traditional solutions for power flow calculations are typically based on constructing and iteratively solving nonlinear equations grounded in physical principles. Commonly employed methods include the Gauss-Seidel method \cite{teng2002modified}, the Newton-Raphson method \cite{chatterjee2017novel}, and their various improvements \cite{huang1994managing, nur2021load}. Although these numerical methods offer high computational accuracy, they are computationally intensive and time-consuming \cite{mohsin2022comparison}. This limitation becomes particularly problematic in large-scale power systems. Moreover, they typically rely on deterministic models, which assume fixed values for all input variables \cite{xu2020probabilistic}. However, with the significantly increasing penetration of renewables in future power systems, the power flow will be inherently uncertain. A potential solution is the Probabilistic power flow (PPF) \cite{pareek2020gaussian}, which provides a probabilistic distribution of possible outcomes rather than a single solution, enabling operators to better anticipate and manage risks. However, the probabilistic power flow model, despite its advantages, inherently remains an analytical method for solving power flow equations. As it ultimately produces an analytical probabilistic model of power flow \cite{jin2023optimal}, the computational complexity and time required can significantly exceed those of traditional power flow calculation methods. 

In recent years, data-driven methods utilizing deep neural networks for power flow calculations \cite{yang2019fast, xiao2023novel, tiwari2024power, xiang2020probabilistic, wang2020automatic} have gained substantial interest due to their strong abilities to model complex nonlinear relationships and extract relevant data features. These approaches leverage large sets of historical power system data to identify patterns that can improve power flow predictions. By shifting the heavy computational load to the offline training phase \cite{yang2019fast}, these data-driven techniques enable rapid power flow predictions during the inference stage through efficient forward propagation, thereby greatly improving calculation speed for online deployment \cite{xiang2020probabilistic}. However, the effectiveness of these methods is heavily dependent on the quality of the training data \cite{tiwari2024power}. These approaches often lack an understanding of the underlying physical principles and characteristics of power systems, such as Kirchhoff's laws and power flow sensitivities. As a result, their predictions can be inconsistent with the actual physical behaviors of the power system, making it challenging to apply these methods broadly in real-world power system operations.

To tackle this issue, the physical-informed neural networks (PINN) is introduced in many existing research \cite{hu2020physics, lei2020data, gao2023physics, chen2023physics, sun2024physics} for power flow calculation. PINNs incorporate the physical laws governing power systems, such as Kirchhoff's laws \cite{hu2020physics} and power flow equations \cite{lei2020data}, directly into the neural network's loss function. This integration ensures that the model's predictions are consistent with known physical principles. However, PINN still faces two significant challenges. Firstly, as the size of the power grid increases, the amount of training data required also escalates dramatically \cite{yang2024probabilistic, lopez2024optimal}, leading to a substantial rise in training efficiency, i.e., training time and cost. Additionally, these methods often struggle with generalization capability \cite{huang2022applications, liu2024voltage}, i.e., their predictive performance may deteriorate when encountering unseen input data. Given the growing uncertainty in power injections caused by the increasing integration of renewable energy sources, the accuracy of these models will be significantly affected. 

The technical challenge addressed in this paper is: how to design an innovative power flow calculation method that can be effectively applied in large-scale power systems with high penetration of renewable energy? The method must not only accelerate training efficiency and reduce time and cost but also ensure generalization, maintaining accuracy even when faced with previously unseen input data. To tackle this challenge, we propose a power flow calculation model based on Bayesian quantum neural networks (BQNN). Firstly, we introduce an advanced quantum deep learning framework that leverages quantum advantages to enhance the deep learning model's ability to capture and represent complex data relationships. Secondly, we employ Bayesian training methods to train the quantum deep learning model, allowing the model to continuously update its understanding of how renewable energy uncertainties impact power flow results through the posterior probability distribution, thereby improving predictive capabilities when dealing with unknown inputs. Lastly, we introduce two metrics to validate the model's capacity to represent data and its generalization ability. Specifically, the contributions of this paper are as follows:

\begin{itemize}
    \item We propose a novel power flow calculation method based on Quantum Neural Networks (QNNs). Specifically, the architecture includes three key elements: Encoding, Ansatz, and Observation, which allow the quantum circuit to operate in a high-dimensional space with improved efficiency, while optimizing its parameters using a similar approach as conventional neural networks. The advantage of this approach lies in its ability to significantly enhance computational efficiency through the use of quantum superposition and entanglement, thereby capturing complex data relationships in large-scale power flow calculation problems.
\end{itemize}

\begin{itemize}
    \item We propose a Bayesian training method for the above QNN to address the challenge of generalizing to unseen power flow data, particularly in the presence of epistemic uncertainties arising from renewable energy sources. This method leverages the posterior distribution of quantum circuit parameters, representing the updated knowledge of power flow uncertainties after observing new data. By incorporating this posterior distribution during both model training and online deployment, the method enables the QNN to capture the impact of renewable generation uncertainties on power flow more effectively.
\end{itemize}

\begin{itemize}
    \item We propose a model evaluation framework focusing on the complexity and generalization ability of the BQNN model. This framework introduces the effective dimension as a measure of model complexity, representing the number of parameters required for accurately capturing the complexity of the data. Additionally, our framework evaluates the generalization ability of the model using the generalization error bound, which quantifies the model’s performance on unseen data under the uncertainties brought by renewable energy.
\end{itemize}

The remainder of this paper is organized as follows. Section 2 formulates the power flow calculation problem. Section 3 introduces the design of BQNN model with Bayesian training method. Section 4 elaborates the model evaluation metrics. Section 5 demonstrates the case study results. Section 6 concludes this paper.

\section{Problem Statement}
Our task is to solve the power flow equations by neural networks (NN). Consider the input $\mathcal{X} := \{\mathbf{P}, \mathbf{Q}\} = \{{P}_i, {Q}_i\}_{i=1}^n$, where ${P}_i$ and ${Q}_i$ denote the active and reactive power injections at bus $i$. The output, voltage magnitudes and phase angles, are denoted as $\mathcal{Y} := \{\mathbf{V}, \mathbf{\phi}\} = \{{V}_i, {\phi}_i\}_{i=1}^n$. Once the output variables are obtained, the power flow can be easily computed by:
\begin{align}
    P_{ij} &= G_{ij} (V_i V_j \cos \phi_{ij}) - B_{ij} V_i \sin \phi_{ij} \label{eq:3.42} \\
    Q_{ij} &= -B_{ij} (V_i^2 - V_i V_j \cos \phi_{ij}) - G_{ij} V_i \sin \phi_{ij} \label{eq:3.43}
\end{align}
where $P_{ij}$ and $Q_{ij}$ denote the active and reactive power flows from bus $i$ to bus $j$, respectively. $G_{ij}$ and $B_{ij}$ are the conductance and susceptance between buses $i$ and $j$. 

The training process of a deep neural network can be viewed as a data fitting problem. The network parameters to be trained are denoted by $\mathbf{\theta} = \{\mathbf{W}, \mathbf{b}\}$. Here, $\mathbf{W}$ represents the weight parameters and $\mathbf{b}$ denotes the bias parameters. The goal is to minimize the distance between the network's output and the actual values by minimizing the loss function. The loss function is defined by the mean squared error:
\begin{align}
    loss = \| \mathcal{Y} - \hat{\mathcal{Y}} \|^2 + \|\mathbf{P}_{out} - \hat{\mathbf{P}}_{out} \|^2 + \| \mathbf{Q}_{out} - \hat{\mathbf{Q}}_{out} \|^2 
\end{align}
where $\mathcal{Y}_{out}$ is the output vector, i.e., voltage magnitude and angle. $\hat{\mathcal{Y}}$ represents the output from the deep neural network.  $\mathbf{P}_{out}$ and $\mathbf{Q}_{out}$ are the actual active and reactive power flows, while $\hat{\mathbf{P}}_{out}$ and $\hat{\mathbf{Q}}_{out}$ are the corresponding estimates from the deep neural network. These variables are all normalized to ensure that each component's importance is comparable. 

As previously mentioned in Section 1, future large-scale power systems with high renewable penetration face two significant challenges in power flow calculation: training efficiency and generalization. Specifically, the training efficiency problem refers to the exponentially increasing number of training samples \( m \) required to achieve the desired training accuracy as the number of nodes \( n \) in the power grid grows. This results in unacceptable training time and cost. The generalization problem arises from the uncertainty of renewables, which means that in practical applications, the input vector \(\mathcal{X} := \{{P}_i, {Q}_i\}_{i=1}^n\) may differ significantly from the training dataset, thereby adversely affecting the accuracy of the predictions.

\section{Bayesian Quantum Neural Network Design}

In this section, we will use the Bayesian quantum neural networks (BQNN) composed by quantum circuits, to replace conventional NN and solve the aforementioned technical problems. We will answer the following two questions: 1) How the BQNN is designed by quantum circuit, and why it can improve the computational efficiency? 2) How the Bayesian training method, which has been successfully implemented in conventional NN, can be implemented in QNN, to mitigate the impact of renewable generations' uncertainty on the accuracy of power flow estimation?

\subsection{General Framework Design}

The overall architecture of the BQNN is depicted in Fig.1. With the following three key elements—Encoding, Ansatz, and Observation—the complex relationship between input and output is represented by the quantum circuit in a high-dimensional space with enhanced expression capability and processed with higher efficiency.

\textbf{\textit{Step 1: Encoding}}
The first step in designing a quantum circuit is encoding the input data onto the qubits to initiate subsequent computations. Generally, the dimensionality of the input features should match the number of qubits. Given that real-world power systems typically have hundreds or even thousands of nodes, the number of available qubits in most quantum computers is limited to less than one hundred. Even when using GPU simulations of quantum circuits, the computational speed significantly decreases when the number of qubits exceeds a few dozen \cite{zhang2020recent}. To tackle this, we employ a hybrid classical-quantum approach to reduce the dimensionality before encoding. The classical neural network component first reduces the high-dimensional input space $\mathbb{R}^{n}$ (where $n$ is the number of nodes in the power network) to a lower-dimensional space $\mathbb{R}^m$ (where $m$ is the number of available qubits). This process is expressed as:
\begin{align}
(\mathbf{P}, \mathbf{Q}) \in \mathbb{R}^{n} \xrightarrow{\text{Classical NN}} (\mathbf{P}', \mathbf{Q}') \in \mathbb{R}^m
\end{align}
where $(\mathbf{P}', \mathbf{Q}')$ represent the reduced-dimensional active and reactive power injections, respectively.

Next, the reduced inputs $(\mathbf{P}', \mathbf{Q}')$ are encoded onto qubits, transforming them into quantum states. We begin by initializing the qubits to an initial state:
\begin{align}
|\psi_0 \rangle = |0\rangle^{\otimes m}
\end{align}
Then, we apply the encoding operation $\mathcal{E}(\mathbf{P}', \mathbf{Q}')$ to obtain the encoded quantum state:
\begin{align}
|\psi_{\mathcal{E}}(\mathbf{P}', \mathbf{Q}')\rangle = \mathcal{E}(\mathbf{P}', \mathbf{Q}') |\psi_0 \rangle
\end{align}
Here, $\mathcal{E}(\mathbf{P}', \mathbf{Q}')$ represents the encoding unitary operation, which includes gates of the form $\exp(-i P'_k H)$ and $\exp(-i Q'_k H)$. These gates perform rotation operations on the qubits, where $P'_k$ and $Q'_k$ are the data to be encoded, $i$ is the imaginary unit, and $H$ denotes the Hamiltonian—an operator that describes the energy of a quantum system and governs its time evolution \cite{peral2024systematic}. Generally, the Hamiltonian $H$ can be any operator that fits the system's requirements, but commonly used ones are the Pauli matrices (e.g., $H = \sigma_x$, $H = \sigma_y$, $H = \sigma_z$), which describe rotations of a single qubit in different directions.

Based on the encoded quantum state, the parameterized quantum circuit $\mathcal{U}_{\bm{\theta}}$ then evolves it into the targeted state:
\begin{align}
|\psi'_{\bm{\theta}}(\mathbf{P}', \mathbf{Q}')\rangle = \mathcal{U}_{\bm{\theta}} |\psi_{\mathcal{E}}(\mathbf{P}', \mathbf{Q}')\rangle
\end{align}
This final state $|\psi'_{\bm{\theta}}(\mathbf{P}', \mathbf{Q}')\rangle$ represents the system within the high-dimensional Hilbert space $\mathcal{H}$ \cite{schuld2019quantum}.

\textbf{Remark 1.} (\textit{How is computational efficiency improved?}) By converting the data to quantum states, the superposition property is leveraged \cite{alchieri2021introduction}. It allows $|\psi'_{\bm{\theta}}(\mathbf{P}', \mathbf{Q}')\rangle$ to exist as a linear combination of all possible basis states simultaneously. This capability significantly enhances computational efficiency, as multiple computations can be performed in parallel. For instance, given an $m$-qubit system, the state $|\psi'_{\bm{\theta}}(\mathbf{P}', \mathbf{Q}')\rangle$ can be expressed as a superposition:
\begin{align}
|\psi'_{\bm{\theta}}(\mathbf{P}', \mathbf{Q}')\rangle = \sum_{i=0}^{2^m-1} c_i |i\rangle
\end{align}
where $c_i$ are complex coefficients and $|i\rangle$ are the computational basis states. This superposition allows the quantum circuit to process $2^m$ states simultaneously, exponentially increasing the amount of information handled compared to classical systems.

\textbf{\textit{Step 2: Ansatz Design}}
An ansatz refers to a parameterized quantum circuit $\mathcal{U}_{\bm{\theta}}$, analogous to the fully connected layers in classical neural networks \cite{alchieri2021introduction}. The design of the ansatz is crucial for the performance of QNNs, as it directly impacts the ability to express complex quantum states and computational efficiency. In this paper, the ansatz structure consists of 2-4 layers, where each layer includes parameterized single-qubit gates and entangling operations to enable the expression and manipulation of complex quantum states. The single-qubit gates used are typically the $R_y$ and $R_z$ rotations, given by the following formulas:
\begin{align}
R_y(\theta) &= \exp\left(-i\frac{\theta}{2}\sigma_y\right) = \begin{pmatrix}
\cos\frac{\theta}{2} & -\sin\frac{\theta}{2} \\
\sin\frac{\theta}{2} & \cos\frac{\theta}{2}
\end{pmatrix} \\
R_z(\theta) &= \exp\left(-i\frac{\theta}{2}\sigma_z\right) = \begin{pmatrix}
\exp\left(-i\frac{\theta}{2}\right) & 0 \\
0 & \exp\left(i\frac{\theta}{2}\right)
\end{pmatrix}
\end{align}
These gates apply rotations around the $y$-axis and $z$-axis of the Bloch sphere, respectively, with $\theta$ being the tunable parameter for each qubit. 

Entanglement \cite{martin2022quantum} among qubits is created using Controlled-NOT (CNOT) gates, which are given by:
\begin{align}
\text{CNOT}_{ij} &= |0\rangle \langle 0|_i \otimes I_j + |1\rangle \langle 1|_i \otimes X_j
\end{align}
where $\text{CNOT}_{ij}$ denotes the CNOT gate with qubit $i$ as the control and qubit $j$ as the target. Here, $|0\rangle \langle 0|_i \otimes I_j$ indicates that if the control qubit $i$ is in the state $|0\rangle$, the target qubit $j$ remains unchanged as denoted by the identity operation $I_j$. Conversely, $|1\rangle \langle 1|_i \otimes X_j$ indicates that if the control qubit $i$ is in the state $|1\rangle$, the target qubit $j$ is flipped by the Pauli-X gate $X_j$.

\textbf{Remark 2.} (\textit{Why is entanglement essential in power flow calculations?}) Entanglement mirrors the physical reality of power balance constraints in the grid. When the power injection $P'_i$ or $Q'_i$ at one node changes, it necessitates compensatory changes in $P'_j$ or $Q'_j$ at other nodes to maintain overall system balance, effectively capturing inter-node dependencies. Additionally, nodal voltage magnitudes and phase angles, as well as active and reactive power flows in transmission lines, exhibit similarly complex entanglement. A fluctuation in one node's voltage or a line's power will influence other nodes' voltages and other lines' power flows.

\textbf{\textit{Step 3: Observation}}
The final step involves measuring the quantum state to compute the model's output, specifically the node voltages $\mathbf{V}$ and phase angles $\mathbf{\phi}$. These outputs are obtained by measuring appropriate observables on the quantum state:
\begin{align}
\mathcal{Y} := \{\mathbf{V}, \mathbf{\phi}\} = f_{\bm{\theta}}(\mathbf{P}', \mathbf{Q}') &= \left\{ \langle \psi'_{\bm{\theta}} | \mathcal{Q}_{V} | \psi'_{\bm{\theta}} \rangle, \ \langle \psi'_{\bm{\theta}} | \mathcal{Q}_{\phi} | \psi'_{\bm{\theta}} \rangle \right\}
\end{align}
Here, the Hermitian operators $\mathcal{Q}_{V}$ and $\mathcal{Q}_{\phi}$ are defined to correspond to the measurements of node voltages and phase angles, respectively. For each node $i$, the voltage $V_i$ and phase angle $\phi_i$ are obtained by measuring specific observables on the qubits: $V_i = \langle \psi'_{\bm{\theta}} | \mathcal{Q}_{V_i} | \psi'_{\bm{\theta}} \rangle$, $\phi_i = \langle \psi'_{\bm{\theta}} | \mathcal{Q}_{\phi_i} | \psi'_{\bm{\theta}} \rangle$. Furthermore, to map the quantum measurements to the desired outputs, a classical post-processing layer can be employed. Specifically, the raw measurement results from the quantum circuit are fed into a classical fully connected neural network layer. This layer processes the quantum measurement data to produce the final estimates of node voltages and phase angles.

\begin{figure*}[!t]
\centering
\includegraphics[width=\textwidth]{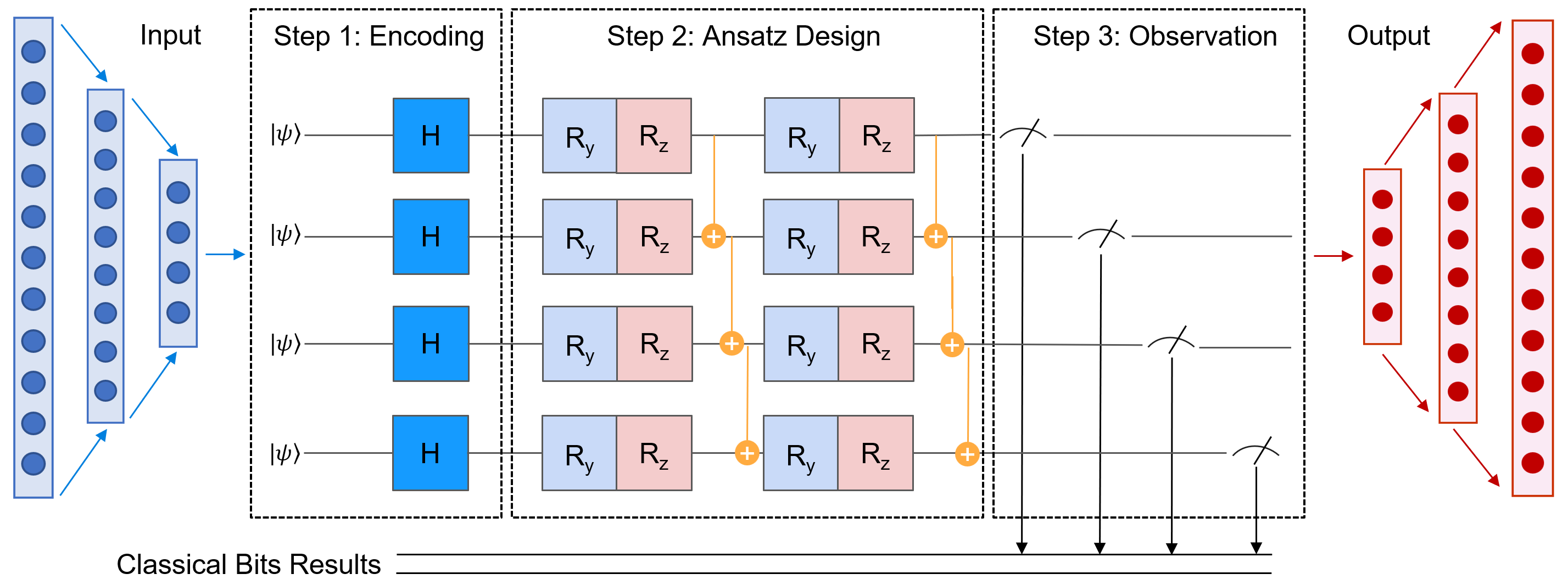}
\caption{Overall Architecture of BQNN Model}
\label{fig_sim}
\end{figure*}

\subsection{Bayesian Training}
In this subsection, we address the generalization problem of Quantum Neural Networks (QNNs) when faced with unseen power flow data caused by epistemic uncertainties in renewable generation. We propose a Bayesian training method that leverages the posterior distribution of model parameters, representing the "updated knowledge" of power flow uncertainties after observing new data. This approach enables capturing the impact of renewable generation uncertainties on power flow, both during model training and after online deployment.

Consider the dataset $\mathcal{D} = \{ (\mathbf{x}_i, \mathbf{y}_i) \}_{i=1}^N$, where $\mathbf{x}_i$ represents the input features, and $\mathbf{y}_i$ represents the corresponding outputs. Specifically, $\mathbf{x}_i = (\mathbf{V}_i, \bm{\phi}_i)$ includes the voltage magnitudes and phase angles at all nodes, and $\mathbf{y}_i = (\mathbf{P}_i, \mathbf{Q}_i)$ represents the active and reactive power flows. The posterior distribution $p(\bm{\theta} | \mathcal{D})$ \cite{nguyen2022bayesian}, representing the distribution of the quantum circuit parameters $\bm{\theta}$ given the data $\mathcal{D}$, can be expressed using Bayes' theorem:
\begin{align}
p(\bm{\theta} | \mathcal{D}) \propto p(\mathcal{D} | \bm{\theta}) \, p(\bm{\theta})
\end{align}
where $p(\mathcal{D} | \bm{\theta})$ is the likelihood function, indicating the probability of observing the data $\mathcal{D}$ given the parameters $\bm{\theta}$, and $p(\bm{\theta})$ is the prior distribution, reflecting our prior belief about the parameters before observing the data. To estimate the posterior distribution $p(\bm{\theta} | \mathcal{D})$, we employ variational inference. We introduce an approximate posterior distribution $q(\bm{\theta}; \bm{\mu})$, parameterized by $\bm{\mu}$, and aim to find the parameters $\bm{\mu}$ that minimize the Kullback-Leibler (KL) divergence between $q(\bm{\theta}; \bm{\mu})$ and the true posterior $p(\bm{\theta} | \mathcal{D})$ \cite{nguyen2022bayesian}:
\begin{align}
\bm{\mu}^* = \arg \min_{\bm{\mu}} \text{KL}\left[ q(\bm{\theta}; \bm{\mu}) \, \| \, p(\bm{\theta} | \mathcal{D}) \right]
\end{align}
Minimizing this KL divergence is equivalent to maximizing the Evidence Lower Bound (ELBO):
\begin{align}
\mathcal{L}(\bm{\mu}) = \mathbb{E}_{q(\bm{\theta}; \bm{\mu})} \left[ \log p(\mathcal{D} | \bm{\theta}) \right] - \text{KL}\left[ q(\bm{\theta}; \bm{\mu}) \, \| \, p(\bm{\theta}) \right]
\end{align}
We optimize $\bm{\mu}$ by maximizing $\mathcal{L}(\bm{\mu})$. A practical approach is to approximate the expectation using Monte Carlo sampling:
\begin{align}
\mathcal{L}(\bm{\mu}) \approx \frac{1}{M} \sum_{k=1}^M \left( \log p(\mathcal{D} | \bm{\theta}^{(k)}) \right) - \text{KL}\left[ q(\bm{\theta}; \bm{\mu}) \, \| \, p(\bm{\theta}) \right]
\end{align}
where $\bm{\theta}^{(k)}$ are samples drawn from the approximate posterior $q(\bm{\theta}; \bm{\mu})$.

After obtaining the approximate posterior distribution $q(\bm{\theta}; \bm{\mu}^*)$, which captures the updated knowledge of the power flow uncertainties, we make predictions in a probabilistic manner. Specifically, we sample multiple parameter sets $\bm{\theta}_1, \bm{\theta}_2, \dots, \bm{\theta}_S$ from the approximate posterior $q(\bm{\theta}; \bm{\mu}^*)$. For a new input $\mathbf{x}_{\text{new}}$, the model makes predictions using each sampled parameter set:
\begin{align}
\hat{\mathbf{y}}_s = f(\mathbf{x}_{\text{new}}; \bm{\theta}_s), \quad s = 1, 2, \dots, S
\end{align}
Each prediction $\hat{\mathbf{y}}_s$ corresponds to a different parameter set $\bm{\theta}_s$, capturing the uncertainty in the model parameters. The final prediction is obtained by averaging the individual predictions: $\hat{\mathbf{y}} = \frac{1}{S} \sum_{s=1}^S \hat{\mathbf{y}}_s$. This Bayesian approach allows the model to account for uncertainty in the predictions, improving generalization to unseen data.

\section{Model Evaluation}
In this section, we focus on evaluating the performance of the model in the context of power flow estimation. For large-scale power systems, accuracy and training efficiency are crucial metrics, typically measured by the root mean square error (RMSE) and the number of training epochs, respectively. Our emphasis, however, is on the model's complexity and generalization ability. Model complexity refers to the number of parameters required for accurate power flow estimation and is quantified using the effective dimension. Generalization ability pertains to the model's accuracy in predicting power flows in unseen scenarios, especially those influenced by the uncertainty of renewable energy outputs, and is measured by the generalization error bound.
 
\subsection{Effective Dimension}
The effective dimension is a key indicator of model complexity. It essentially represents the number of parameters needed for the model to accurately capture the complexity of the data. A higher effective dimension indicates a model's capacity to capture complex data features but may also heighten the risk of overfitting. It is mathematically defined as \cite{nguyen2022bayesian}:
\begin{align}
d_{\gamma,n}(g_{\bm{\theta}}) &= 2 \frac{\log \left( \dfrac{1}{V_{\Theta}} \int_{\Theta} \sqrt{\det \left( \mathbf{I}_d + \zeta \mathbf{F}(\bm{\theta}) \right)} \, d\bm{\theta} \right)}{\log \zeta}
\end{align}
where $\zeta = \dfrac{\gamma n}{2\pi \log n}$, $\Theta \subset \mathbb{R}^d$ is the parameter space with $d$ being the number of parameters in $\bm{\theta}$, $\gamma \in (0,1]$ is a regularization parameter, and $n$ is the number of samples. The term $V_{\Theta} = \int_{\Theta} d\bm{\theta}$ represents the volume of the parameter space, and $\mathbf{I}_d$ is the $d \times d$ identity matrix. $\mathbf{F}(\bm{\theta})$ denotes the Fisher information matrix, which is generally intractable.

To practically compute its estimated value $\hat{\mathbf{F}}_n(\bm{\theta})$, we use the following method. Assume that after model training, we have obtained the parameter distribution $p(\mathbf{y}|\mathbf{x}; \bm{\theta}) p(\mathbf{x})$ that indicates the probabilistic power flow. Given a finite sample size $n$, the empirical Fisher information matrix $\hat{\mathbf{F}}_n(\bm{\theta})$ can be estimated using observed data samples $(\mathbf{x}_i, \mathbf{y}_i)$:
\begin{align}
\hat{\mathbf{F}}_n(\bm{\theta}) = \frac{1}{n} \sum_{i=1}^{n} \left( \nabla_{\bm{\theta}} \log p(\mathbf{y}_i | \mathbf{x}_i; \bm{\theta}) \right) \left( \nabla_{\bm{\theta}} \log p(\mathbf{y}_i | \mathbf{x}_i; \bm{\theta}) \right)^\top
\end{align}
Here, $\nabla_{\bm{\theta}} \log p(\mathbf{y}_i | \mathbf{x}_i; \bm{\theta})$ denotes the gradient of the log-likelihood with respect to the parameters $\bm{\theta}$.

To construct the likelihood function for a given dataset \( \mathcal{D} = \{ (\mathbf{x}_i, \mathbf{y}_i) \}_{i=1}^n \), we model the discrepancy between the model output \( f_{\bm{\theta}}(\mathbf{x}_i) \) and the observed value \( \mathbf{y}_i \) as a Gaussian distribution. Recall that the output of a quantum neural network is generally represented by the expectation value of an observable measured on a quantum state. Suppose we have a quantum circuit \( \mathcal{U}_{\bm{\theta}} \) and an input quantum state \( \psi \); the output is the expectation value of an observable \( \mathcal{Q} \), denoted as \( f_{\bm{\theta}}(\mathbf{x}_i) \). Assuming the errors follow a multivariate normal distribution with mean \( f_{\bm{\theta}}(\mathbf{x}_i) \) and covariance matrix \( \sigma^2 \mathbf{I}_k \), where \( k \) is the dimension of \( \mathbf{y}_i \), the likelihood function can be written as:
\begin{align}
p(\mathbf{y}_i | \mathbf{x}_i; \bm{\theta}) = \frac{1}{(2\pi \sigma^2)^{k/2}} \exp \left( -\frac{\left\| \mathbf{y}_i - f_{\bm{\theta}}(\mathbf{x}_i) \right\|^2}{2 \sigma^2} \right)
\end{align}
where \( \| \cdot \| \) denotes the Euclidean norm. The log-likelihood function is then:
\begin{align}
\log p(\mathbf{y}_i | \mathbf{x}_i; \bm{\theta}) = -\frac{k}{2} \log(2\pi \sigma^2) - \frac{\left\| \mathbf{y}_i - f_{\bm{\theta}}(\mathbf{x}_i) \right\|^2}{2 \sigma^2}
\end{align}
To calculate the gradient of the log-likelihood function with respect to the parameter \( \bm{\theta} \), we have:
\begin{align}
\nabla_{\bm{\theta}} \log p(\mathbf{y}_i | \mathbf{x}_i; \bm{\theta}) = \frac{1}{\sigma^2} \left( \mathbf{y}_i - f_{\bm{\theta}}(\mathbf{x}_i) \right)^\top \nabla_{\bm{\theta}} f_{\bm{\theta}}(\mathbf{x}_i)
\end{align}
where \( \nabla_{\bm{\theta}} f_{\bm{\theta}}(\mathbf{x}_i) \) is the Jacobian matrix of the model output with respect to the parameters \( \bm{\theta} \).

\textbf{Remark 3. }(\textit{Backpropagation in Quantum Circuits}) Suppose that we define the loss function \( \mathcal{L}(\bm{\theta}) \) of our BQNN as the mean squared error loss:
\begin{align}
\mathcal{L}(\bm{\theta}) = \frac{1}{n} \sum_{i=1}^{n} \left\| f_{\bm{\theta}}(\mathbf{x}_i) - \mathbf{y}_i \right\|^2
\end{align}
To update the parameter \( \bm{\theta} \), we need to calculate the gradient of the loss function with respect to each parameter \( \theta_j \). In the quantum neural network, this gradient can be calculated using the parameter shift rule \cite{kamruzzaman2020quantum}. First, we compute the gradient of the model output with respect to the parameters:
\begin{align}
\nabla_{\theta_j} f_{\bm{\theta}}(\mathbf{x}_i) = \frac{1}{2} \left( f_{\bm{\theta} + \tfrac{\pi}{2} e_j}(\mathbf{x}_i) - f_{\bm{\theta} - \tfrac{\pi}{2} e_j}(\mathbf{x}_i) \right)
\end{align}
where \( e_j \) is the unit vector in the direction of the \( j \)-th parameter. Then, we compute the gradient of the loss function:
\begin{align}
\frac{\partial \mathcal{L}}{\partial \theta_j} = \frac{2}{n} \sum_{i=1}^{n} \left( f_{\bm{\theta}}(\mathbf{x}_i) - \mathbf{y}_i \right)^\top \nabla_{\theta_j} f_{\bm{\theta}}(\mathbf{x}_i)
\end{align}
Finally, the parameters can be updated using gradient descent as follows:
\begin{align}
\theta_j \leftarrow \theta_j - \eta \frac{\partial \mathcal{L}}{\partial \theta_j}
\end{align}
where \( \eta \) is the learning rate.

\subsection{Generalization Error Bound}

In machine learning, the generalization error is crucial as it measures a model's performance on unseen data. It can be estimated based on the Vapnik-Chervonenkis (VC) dimension \cite{abu1989vapnik}, which quantifies the capacity of a hypothesis class. The VC dimension of a model class is the largest number of points that can be shattered by the model, where "shattered" means that the model can classify the points in all possible ways.

For our proposed BQNN model \( f_{\bm{\theta}} \), the generalization error bound \cite{suzuki2018fast} is given by the formula:
\begin{align}
R(f_{\bm{\theta}}) \leq \hat{R}(f_{\bm{\theta}}) + \sqrt{\dfrac{h \left( \log \dfrac{n}{h} + 1 \right) + \log \dfrac{1}{\delta}}{n}}
\end{align}
where \( R(f_{\bm{\theta}}) \) represents the true risk or generalization error of the model \( f_{\bm{\theta}} \), and \( \hat{R}(f_{\bm{\theta}}) \) is the empirical risk, which measures the model's error on the training data. The term \( h \) is the VC dimension of the model class, \( n \) denotes the number of training samples, and \( \delta \in (0,1) \) is the confidence level. The inequality holds with probability at least \( 1 - \delta \).

To apply this formula, one must first determine the VC dimension \( h \) of the model class. Estimating the VC dimension of a hybrid model combining classical deep learning and quantum circuits involves considering the parameters of both components. For the classical neural network part, the VC dimension is typically proportional to the total number of weights and biases, denoted as \( P_{\text{classical}} \) \cite{bartlett2003vapnik}. For the quantum neural network part, estimating the VC dimension is more challenging due to the quantum nature of computations. However, as an approximation, the VC dimension of the quantum component can be related to the number of quantum circuit parameters \( P_{\text{quantum}} \) \cite{suzuki2018fast}. Therefore, the overall VC dimension of the hybrid model can be estimated by combining the VC dimensions of the classical and quantum components:
\begin{align}
h_{\text{hybrid}} \approx P_{\text{classical}} + P_{\text{quantum}}
\end{align}
Next, the empirical risk \( \hat{R}(f_{\bm{\theta}}) \) is calculated by evaluating the model's performance over the training data:
\begin{align}
\hat{R}(f_{\bm{\theta}}) = \frac{1}{n} \sum_{i=1}^{n} \mathcal{L}\left( f_{\bm{\theta}}(\mathbf{x}_i), \mathbf{y}_i \right)
\end{align}
where \( \mathcal{L} \) is the loss function measuring the discrepancy between the predicted outputs and the true outputs. In this context, a common choice for \( \mathcal{L} \) is the mean squared error (MSE) loss:
\begin{align}
\mathcal{L}\left( f_{\bm{\theta}}(\mathbf{x}_i), \mathbf{y}_i \right) = \left\| f_{\bm{\theta}}(\mathbf{x}_i) - \mathbf{y}_i \right\|^2
\end{align}
Here, \( \| \cdot \| \) denotes the Euclidean norm. With the number of samples \( n \) and the desired confidence level \( \delta \) specified, the complexity term can be calculated. This term incorporates both the model's complexity and the amount of training data, ensuring the model is neither too simple (risking underfitting) nor too complex (risking overfitting). By combining the empirical risk and the complexity term, the generalization error \( R(f_{\bm{\theta}}) \) can be bounded, providing insights into the model's expected performance on unseen data.

\section{Case Study}
\subsection{Basic Settings and Test Systems}
In this section, we present a comprehensive case study to evaluate the performance of BayesianQNNs in power flow calculations under various scenarios. We consider the following scenarios based on the IEEE standard test systems: 
\begin{itemize}
    \item \textbf{1a}:  6-bus system with no renewables.
    \item \textbf{1b}:  6-bus system with 20\% renewable penetration.
    \item \textbf{1c}:  6-bus system with 50\% renewable penetration.
    \item \textbf{2a}:  30-bus system with no renewable generation.
    \item \textbf{2b}:  30-bus system with 20\% renewable penetration.
    \item \textbf{2c}:  30-bus system with 50\% renewable penetration.
    \item \textbf{3a}:  118-bus system with no renewable generation.
    \item \textbf{3b}:  118-bus system with 20\% renewable penetration.
    \item \textbf{3c}:  118-bus system with 50\% renewable penetration.
\end{itemize}

\begin{table*}[h!]
\centering
\caption{Comparison of Different Algorithms' Accuracy across Various Scenarios  (With Noise)}
\begin{tabular}{ccccccccccccccc}
\hline
\multirow{2}{*}{Scenario} & \multicolumn{3}{c}{Bayesian QNN} & \multicolumn{3}{c}{QNN} & \multicolumn{3}{c}{Bayesian NN} & \multicolumn{3}{c}{MLP} \\
 & $V_{\text{mse}}$ & $\phi_{\text{dev}}$ & $P_{\text{dev}}$ & $V_{\text{mse}}$ & $\phi_{\text{dev}}$ & $P_{\text{dev}}$ & $V_{\text{mse}}$ & $\phi_{\text{dev}}$ & $P_{\text{dev}}$ & $V_{\text{mse}}$ & $\phi_{\text{dev}}$ & $P_{\text{dev}}$ \\
\hline
1a & 1.39 (12.38) & 0.0 (0.18) & 0.0 (0.14) & 2.75 (11.97) & 0.0 (0.20) & 0.0 (0.18) & 2.53 & 0.0 & 0.0 & 2.72 & 0.0 & 0.0 \\
1b & 1.89 (14.56) & 0.04 (0.21) & 0.03 (0.19) & 2.13 (14.13) & 0.09 (0.24) & 0.07 (0.23) & 2.06 & 0.06 & 0.07 & 2.99 & 0.14 & 0.12 \\
1c & 3.18 (22.99) & 0.11 (0.24) & 0.10 (0.22) & 3.32 (23.17) & 0.19 (0.28) & 0.21 (0.19) & 3.25 & 0.14 & 0.15 & 3.31 & 0.22 & 0.20 \\
2a & 1.41 (15.42) & 0.0 (0.21) & 0.0 (0.18) & 2.73 (18.15) & 0.0 (0.17) & 0.0 (0.16) & 2.56 & 0.0 & 0.0 & 2.78 & 0.0 & 0.0 \\
2b & 1.96 (19.59) & 0.06 (0.27) & 0.08 (0.29) & 2.17 (22.14) & 0.15 (0.29) & 0.14 (0.27) & 2.15 & 0.11 & 0.13 & 3.21 & 0.24 & 0.21 \\
2c & 3.37 (21.92) & 0.21 (0.31) & 0.19 (0.27) & 3.59 (23.99) & 0.28 (0.33) & 0.30 (0.39) & 3.46 & 0.27 & 0.29 & 3.52 & 0.33 & 0.38 \\
3a & 1.38 (14.29) & 0.0 (0.32) & 0.0 (0.41) & 2.73 (19.17) & 0.0 (0.26) & 0.0 (0.21) & 2.55 & 0.0 & 0.0 & 2.79 & 0.0 & 0.0 \\
3b & 2.24 (19.88) & 0.13 (0.36) & 0.15 (0.39) & 2.36 (19.23) & 0.14 (0.28) & 0.18 (0.29) & 2.49 & 0.17 & 0.16 & 3.26 & 0.32 & 0.35 \\
3c & 3.78 (25.34) & 0.23 (0.39) & 0.21 (0.44) & 4.12 (24.67) & 0.32 (0.47) & 0.35 (0.49) & 3.99 & 0.30 & 0.31 & 3.59 & 0.41 & 0.43 \\
\hline
\end{tabular}
\end{table*}

We utilize standard IEEE 6-bus, 30-bus, and 118-bus systems, integrating wind and PV with penetration levels of 20\% and 50\% . Power flow data is generated using MATPOWER. The data generation process involves selecting daily load data with a 15-minute resolution and wind/PV output data over one year, followed by economic dispatch and power flow calculations. For the IEEE 6-bus, 30-bus, and 118-bus systems, we randomly select 1,000, 2,000, and 3,000 data points, and we 4 qubits in both BayesianQNN and QNN respectively. Each dataset is split into 60\% for training and 40\% for testing. The computing is performed using NVIDIA RTX 3090 GPU. To comprehensively evaluate the performance, we compare BayesianQNN versus QNN to assess the impact of Bayesian training on model performance; BayesianQNN versus BayesianNN to evaluate the influence of quantum computing on model performance; and BayesianQNN, QNN, and BayesianNN versus MLP to compare with a baseline algorithm and assess overall performance improvements. 

\subsection{Comparison of Accuracy}
The experimental results in Table 1 demonstrate that BayesianQNN consistently outperforms QNN, BayesianNN, and MLP across various scenarios, with its performance advantage becoming more pronounced as the network size and renewable energy penetration increase. In Scenario 1a, BayesianQNN achieves a normalized mean squared error ($V_{\text{mse}}$) of 1.39, which is 49.45\% lower than QNN, 45.06\% lower than BayesianNN, and 48.90\% lower than MLP. However, as renewable energy penetration increases in Scenario 1b and Scenario 1c, the $V_{\text{mse}}$ of BayesianQNN increases to 1.89 and 3.18. Despite this increase, BayesianQNN still outperforms the other algorithms, showing an 11.27\% and 4.22\% lower $V_{\text{mse}}$ than QNN in scenarios 1b and 1c. In the largest system (Scenario 3a), BayesianQNN achieves a $V_{\text{mse}}$ of 1.38, which is 49.45\% lower than QNN, 45.88\% lower than BayesianNN, and 50.54\% lower than MLP. As the renewable penetration increases in Scenario 3b and 3c, BayesianQNN's $V_{\text{mse}}$ increases to 2.24 and 3.78, respectively. Even in these more challenging scenarios, BayesianQNN shows a 5.08\% and 8.25\% lower $V_{\text{mse}}$ than QNN in scenarios 3b and 3c. Furthermore, BayesianQNN shows a significant reduction in the probability of voltage phase angle prediction errors exceeding 0.05 rad ($\phi_{\text{dev}}$) and active power flow prediction errors exceeding 5 MW ($P_{\text{dev}}$) compared to the other models. 

We further incorporated noise into the Pennylane quantum simulator to evaluate the performance of BayesianQNN and QNN. We introduced three types of noise: (1) bit-flip noise, which flips the state of a qubit with a probability of 0.1; (2) phase-flip noise, which flips the phase of a qubit with a probability of 0.1; and (3) depolarizing noise, which fully randomizes the quantum state with a probability of 0.1. These parameter settings were chosen based on the fact that real device noise is typically low, likely ranging from 0 to 0.1. The results of these tests are presented in parentheses in Table 1. Even with the relatively low noise we introduced, it already had a nearly catastrophic effect on both quantum deep learning models. The mean square error (MSE) of the voltage increased by approximately 7 to 10 times, and the probability of voltage phase angles and line power prediction errors exceeding acceptable limits increased by an average of 3.1 times. This result indicates that the success of quantum deep learning algorithms will depend heavily on advances in quantum computer hardware and  error correction technologies. 

\subsection{Comparison of Computation Speed}
In this subsection, we evaluated the speed by determining the number of epochs required to achieve a specific accuracy threshold: maintaining the voltage phase angle error within 0.05 rad and the active power flow error within 5 MW with a probability exceeding 95\%. The results are summarized in Table 2.

In scenarios without renewable energy (1a, 2a, 3a), all four algorithms manage to achieve the targeted accuracy requirement. In these cases, the required number of epochs increases significantly with the number of nodes in the power system. For example, in scenario 1a, Bayesian QNN requires 298 epochs, while QNN requires 38\% more epochs, Bayesian NN requires 22\% more, and MLP requires 48\% more. Bayesian QNN consistently demonstrates the fastest computation speed among the four algorithms, with QNN following closely behind. In scenarios with renewable energy penetration, as the penetration rate increases, so does the uncertainty in power flow, resulting in increased training difficulty and the number of epochs required. However, these algorithms face challenges in the 118-node system scenarios (3b, 3c). This indicates that, despite the efficiency of Bayesian training in handling uncertainty, the complexity of large-scale power systems with high renewable energy penetration presents significant challenges.

However, it is important to note that even though the Bayesian QNN may require fewer episodes, its absolute computation time is still significantly long. Each epoch takes approximately 6-20 minutes, which is far longer than the computation time required for traditional deep learning (several seconds). This is because the quantum simulator requires substantial time to handle quantum entangled and superposition states, as well as to update the quantum circuit parameters. If model training were conducted on a real quantum computer, the required time would likely be significantly reduced. 

\begin{table*}[h!]
\centering
\caption{Comparison of Different Algorithms' Computation Speed across Various Scenarios}
\begin{tabular}{ccccccccccccccc}
\hline
\multirow{2}{*}{Scenario} & \multicolumn{3}{c}{Bayesian QNN} & \multicolumn{3}{c}{QNN} & \multicolumn{3}{c}{Bayesian NN} & \multicolumn{3}{c}{MLP} \\
 & Epoch & $V_{\text{mse}}$ & $\phi_{\text{dev}}$ and $P_{\text{dev}}$ & Epoch & $V_{\text{mse}}$ & $\phi_{\text{dev}}$ and $P_{\text{dev}}$ & Epoch & $V_{\text{mse}}$ & $\phi_{\text{dev}}$ and $P_{\text{dev}}$ & Epoch & $V_{\text{mse}}$ & $\phi_{\text{dev}}$ and $P_{\text{dev}}$ \\
\hline
1a & 298 & 1.42 & $\leq 0.05$ & 412 & 2.83 & $\leq 0.05$ & 363 & 2.59 & $\leq 0.05$ & 443 & 2.75 & $\leq 0.05$ \\
1b & 917 & 1.86 & $\leq 0.05$ & 1615 & 2.99 & $\leq 0.05$ & 1903 & 2.73 & $\leq 0.05$ & 2123 & 3.11 & $\leq 0.05$ \\
1c & 2319 & 3.45 & $\leq 0.05$ & 2725 & 3.98 & $\leq 0.05$ & 3267 & 3.72 & $\leq 0.05$ & 3653 & 4.26 & $\leq 0.05$ \\
2a & 1046 & 1.43 & $\leq 0.05$ & 1621 & 2.76 & $\leq 0.05$ & 1447 & 2.59 & $\leq 0.05$ & 1698 & 2.79 & $\leq 0.05$ \\
2b & 2903 & 1.98 & $\leq 0.05$ & 3512 & 2.28 & $\leq 0.05$ & 3814 & 2.59 & $\leq 0.05$ & 4132 & 2.82 & $\leq 0.05$ \\
2c & 3103 & 1.42 & $\leq 0.05$ & 3998 & 3.61 & $\leq 0.05$ & 3865 & 3.53 & $\leq 0.05$ & 4828 & 3.64 & $\leq 0.05$ \\
3a & 2587 & 1.40 & $\leq 0.05$ & 3025 & 2.82 & $\leq 0.05$ & 3442 & 2.58 & $\leq 0.05$ & 3516 & 2.83 & $\leq 0.05$ \\
3b & 3621 & 2.31 & $\leq 0.05$ & N.A. & N.A. & $\leq 0.05$ & N.A. & N.A. & $\leq 0.05$ & N.A. & N.A. & $\leq 0.05$ & \\
3c & N.A. & N.A. & $\leq 0.05$ & N.A. & N.A. & $\leq 0.05$ & N.A. & N.A. & $\leq 0.05$ & N.A. & N.A. & $\leq 0.05$ & \\
\hline
\end{tabular}
\end{table*}

\subsection{Comparison of Model Complexity}
In this subsection, we evaluate the model complexity of different algorithms using effective dimension. For scenarios 1, 2, and 3, we fed 1,000, 5,000, and 30,000 sample inputs, respectively, into four algorithm models, training them for 500, 1,000, and 3,000 epochs. We then calculated the effective dimension of the models' parameters using the formula from Section 4A. The results are depicted in Fig.2. 

\begin{figure*}[!t]
\centering
\includegraphics[width=\textwidth]{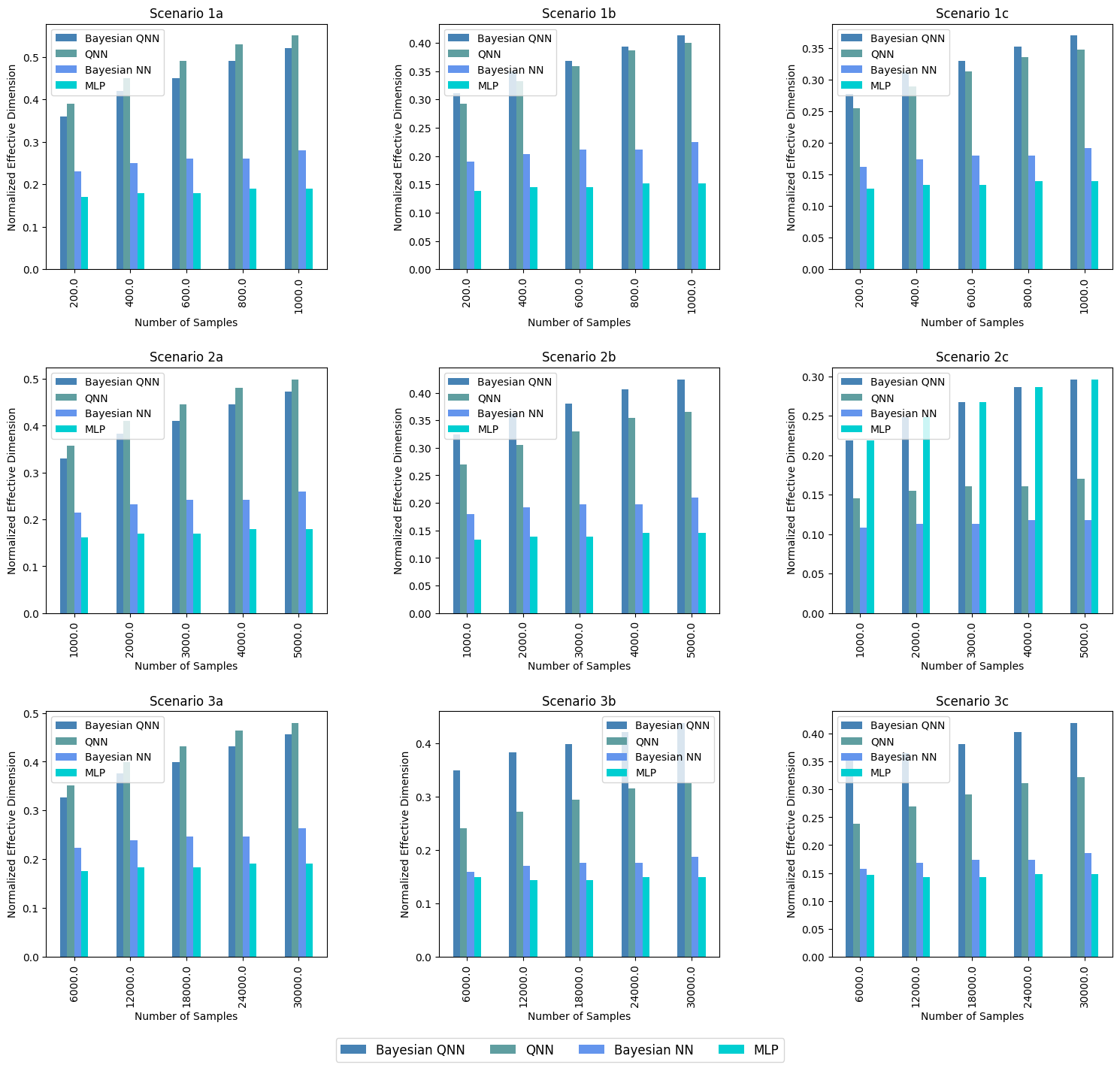}
\caption{Effective Dimension of Different Algorithms}
\label{fig_sim}
\end{figure*}

\begin{figure*}[!t]
\centering
\includegraphics[width=\textwidth]{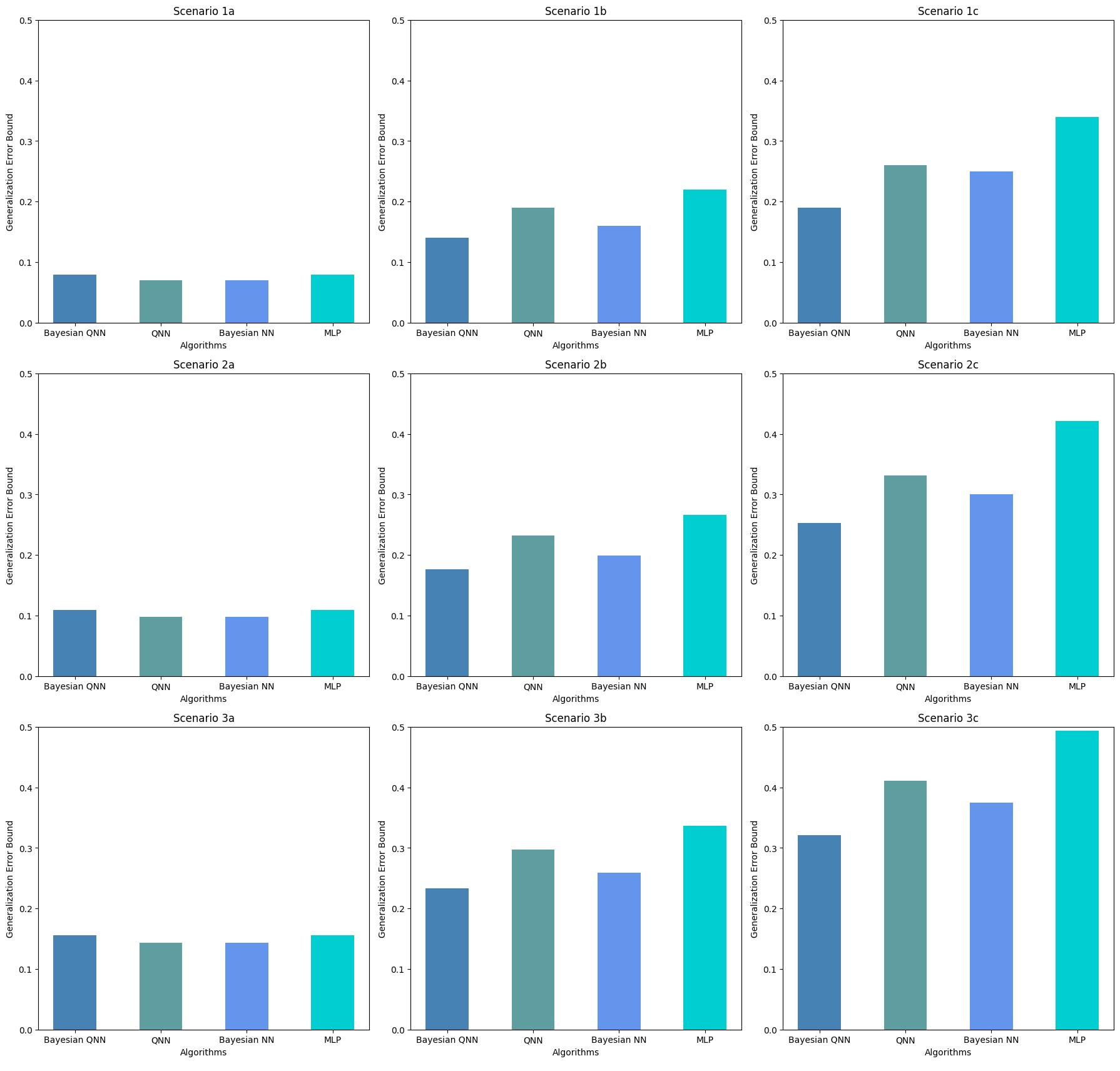}
\caption{Generalization Error Bound of Different Algorithms}
\label{fig_sim}
\end{figure*}

In each scenario, the effective dimension of all algorithms showed an increasing trend with the rise in sample size. For instance, in scenario 1a, Bayesian QNN's effective dimension increased by 44.4\% from 0.36 to 0.52 as the number of samples increased from 200 to 1000. This indicates that an increase in training samples enhances the ability of all algorithms to capture the complex patterns of the power flow equations. In scenarios without the uncertainties introduced by renewable energy generation (1a, 2a, 3a), the effective dimension of QNN was slightly higher than that of Bayesian QNN. This suggests that in the absence of uncertainty, the Bayesian training method does not significantly enhance the model's ability to capture the characteristics of the power flow data. However, it also implies that Bayesian training helps avoid potential overfitting issues. In scenarios with the uncertainties, the effective dimension of Bayesian QNN was significantly higher than that of QNN and other algorithms. For example, in scenario 3c, Bayesian QNN's effective dimension reached 0.456 with 30000 samples, which was 8.6\% higher than QNN's 0.419. This indicates that in these scenarios, the complexity of the data representation is much higher than in scenarios without uncertainty, and only Bayesian QNN is capable of adapting well to this increased complexity. Finally, as the network size increases, the effective dimension of all algorithms except Bayesian QNN tends to decrease. Bayesian QNN, however, maintains a high effective dimension even as the network size grows. Remarkably, in scenario 3b, its effective dimension reached 0.437 with 30000 samples, which is higher than scenario 2c's Bayesian QNN effective dimension of 0.413 with 5000 samples. This demonstrates that with a sufficient number of samples, Bayesian QNN has the potential to continuously capture the data characteristics in large-scale power grid flow calculations.

\subsection{Comparison of Generalization Capability}
In this section, we evaluate the generalization error bounds of various algorithm models to assess their generalization performance. Specifically, we use the formula provided in Section 4B, assuming a confidence parameter of 0.05. Figure 3 illustrates the generalization error results for different algorithms. Generally, a generalization error bound of less than 0.1 indicates that the model has good generalization capability on unseen data \cite{cao2020generalization}. A generalization error bound greater than 0.3 implies that the model performs poorly on unseen data.

In scenarios without renewable energy uncertainty, the generalization error bounds for all four algorithms are mostly within 0.1. Additionally, as the network scale increases, the generalization error bound does not show a significant upward trend from scenario 1a to scenarios 2a and 3a, suggesting that the increase in network size does not severely impact the generalization performance of the algorithms. In scenarios with renewable energy uncertainty, there is a noticeable increase in the generalization error bounds for all four algorithms. Particularly in scenario 3c, the generalization error bounds for QNN and MLP are significantly high, with MLP's reaching up to 0.5, indicating that in these more complex environments, only the Bayesian QNN manages to maintain relatively acceptable generalization performance.

\section{Conclusion}
In this paper, we propose and evaluate a novel BQNN model for power flow calculation, particularly suited for large-scale power systems with high renewable energy penetration. Through comprehensive case studies under various scenarios, we demonstrate that BQNNs consistently outperform conventional algorithms like QNN, Bayesian NN, and MLP, particularly as the network size and renewable energy uncertainty increase. BQNNs achieve lower errors and superior generalization performance due to their ability to incorporate prior knowledge and update predictions based on new data. The evaluation shows that BQNNs maintain high computational efficiency and accuracy, even under the most challenging conditions, highlighting their potential for real-world applications in complex power systems with uncertainties. 

However, practical test also demonstrates the slow processing time and potential errors due to quantum noise, which is constrained by the quantum hardware development.  In future work, we plan to test the performance of quantum deep learning models on actual quantum computers for power flow calculations. We will also develop robust algorithms with better performance under the risk of quantum error, mitigating potential threats while applying to real-world power systems.

\bibliographystyle{IEEEtran}

\bibliography{IEEEabrv,Bibliography}

\newpage
\vfill

\end{document}